\documentstyle[12pt,html,psfig]{article}

\def\reference{\parskip 0pt\par\noindent\hangindent 0.5 truecm}
\newcommand{\kms}{\mbox{\,km\,s$^{-1}$}}
\newcommand{\kpc}{\mbox{\,h$^{-1}$\,kpc}}
\newcommand{\Mpc}{\mbox{\,h$^{-1}$\,Mpc}}
\newcommand{\cubicMpc}{\mbox{\,h$^{-3}$\,Mpc$^3$}}
\newcommand{\invMpc}{\mbox{\,h\,Mpc$^{-1}$}}
\newcommand{\degree}{\mbox{$^{\circ}$}}
\newcommand{\Msol}{\mbox{\,M$_{\odot}$}}
\newcommand{\htmlink}[2]{\htmladdnormallink{#1}{#2}}
\newcommand{\proclink}[1]{\htmladdnormallink{[#1]}
               {http://www.mso.anu.edu.au/DunkIsland/Proceedings/#1}}
\newcommand{\adslink}[2]{\htmladdnormallink{#1}
               {http://adsabs.harvard.edu/cgi-bin/nph-bib_query?bibcode=#2}}
\newcommand{\astroph}[2]{\htmladdnormallink{#1}
               {http://xxx.lanl.gov/abs/astro-ph/#2}}

\begin{document}

\title{Redshift Surveys and Cosmology}

\author{Matthew Colless $^{1}$
}

\date{}

\maketitle

{\center
$^1$ Research School of Astronomy \& Astrophysics, The Australian
National University, \\ Mount Stromlo Observatory, Weston Creek, ACT 2611,
Australia \\ colless@mso.anu.edu.au\\[3mm]
}

\begin{abstract}
Redshift surveys are one of the prime tools of observational cosmology.
Imaging surveys of the whole sky are now available at a wide range of
wavelengths, and provide a basis for the new generation of massive
redshift surveys currently in progress. The very large datasets produced
by these surveys call for new and sophisticated approaches to the
analysis of large-scale structure and the galaxy population. These
issues, and some preliminary results from the new redshift surveys, were
discussed at the second Coral Sea Cosmology Conference, held at Dunk
Island on 24--28 August 1999. This is a summary of the conference; the
full conference proceedings are on the WWW at
\htmlink{http://www.mso.anu.edu.au/DunkIsland/Proceedings}
{http://www.mso.anu.edu.au/DunkIsland/Proceedings}.
\end{abstract}


\bigskip

\section{Introduction}

The second Coral Sea Cosmology Conference, on {\it Redshift Surveys and
Cosmology}, takes place in the midst of a burgeoning of massive redshift
surveys made possible by highly-parallel spectrographs and/or dedicated
telescopes. These redshift surveys stand on the shoulders of even more
massive (in terabyte terms) imaging surveys at a variety of wavelengths,
which provide complete, homogeneous and complementary ways of viewing
the universe and selecting samples for follow-up spectroscopy. The
wealth of detail produced by these surveys is provoking more
sophisticated approaches to data reduction, analysis and archiving,
which offer to both improve the primary science coming out of the
surveys and provide a rich resource for astronomers to mine in the
coming decade.

The outline of this conference summary is as follows: sections~2 and~3
report on the status of redshift surveys, both in progress and planned,
which are mapping the local and high-redshift universe; section~4
summarises some of the preliminary results on large-scale structure
derived from these and other recent surveys; section~5 reports new
results on the local galaxy population and its evolution out to high
redshift; section~6 looks to the future of redshift surveys as tools for
cosmology. All references in [..] are to papers in the WWW proceedings
of the conference, available at
\htmladdnormallink{http://www.mso.anu.edu.au/DunkIsland/Proceedings}
{http://www.mso.anu.edu.au/DunkIsland/Proceedings}.

\section{Redshift Surveys of the Local Universe}

\begin{figure}
\begin{center}
\parbox{\textwidth}{\psfig{file=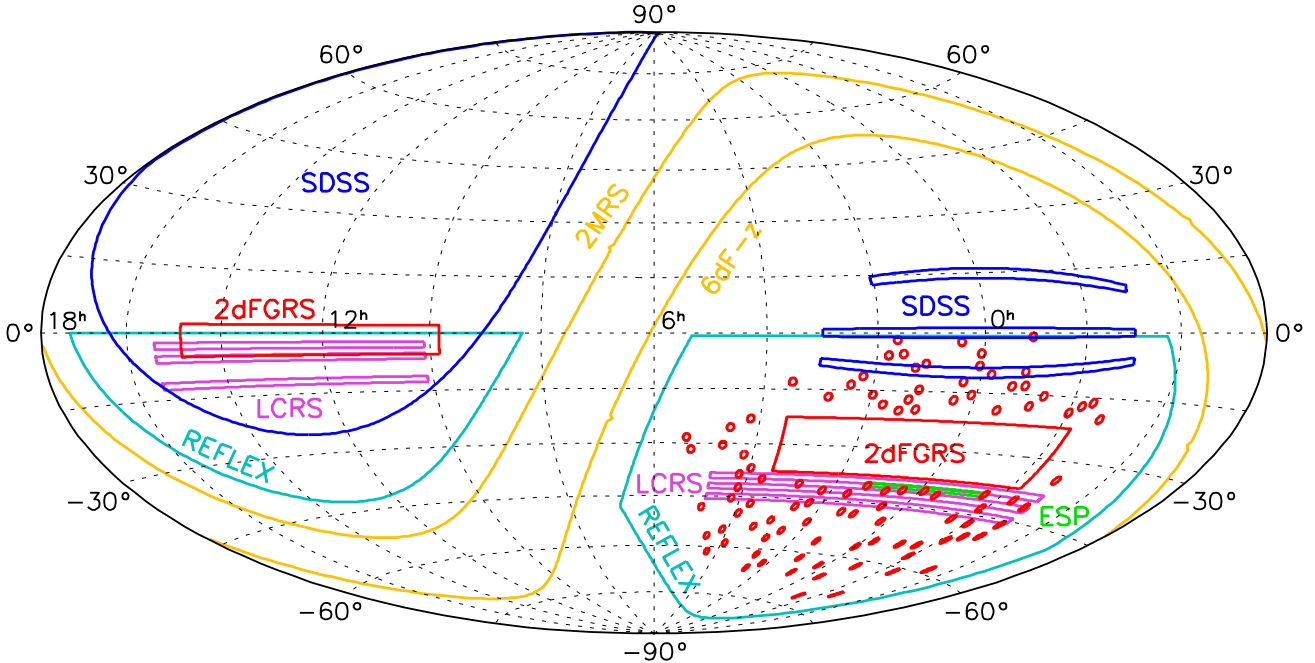,width=\textwidth}}
\caption{The areas of the sky covered by various surveys mentioned in
the text.}
\label{fig:onsky}
\end{center}
\end{figure}

The Two-Degree Field facility (\htmlink{2dF}{http://www.aao.gov.au/2df})
at the Anglo-Australian Telescope is a 400-fibre optical spectrograph
with a 2\degree-diameter field of view. \proclink{Lewis} described how
2dF's fast robotic fibre-positioner and two focal planes makes it
ideally suited for massive redshift surveys. The 2dF Galaxy Redshift
Survey (\htmlink{2dFGRS}{http://www.mso.anu.edu.au/~colless/2dF}) has
now measured redshifts for more than 50,000 galaxies, making it the
single largest redshift survey to date. \proclink{Colless} gave an
overview of the survey's status, reporting that, as of August 1999, 227
of the 1093 survey fields have been observed, yielding redshifts for
53,192 objects, of which 50,180 are galaxies (the remainder are mostly
stars plus a handful of QSOs). The redshift completeness of the survey
is currently 91\% and rising, as data quality and reduction methods are
refined. The 2dFGRS will eventually provide redshifts and spectral
information for 250,000 galaxies with $b_J$$<$19.45
(extinction-corrected) over nearly 2000~sq.deg (\adslink{Colless
1998}{1998wfsc.conf...77}, \astroph{1999}{9804079}).
Figure~\ref{fig:onsky} shows the sky coverage of the 2dFGRS and a number
of the other surveys mentioned here. Although the survey is at present
only 20\% complete, it is expected to be finished by the end of 2001.

\begin{figure}
\begin{center}
\parbox{0.85\textwidth}{\psfig{file=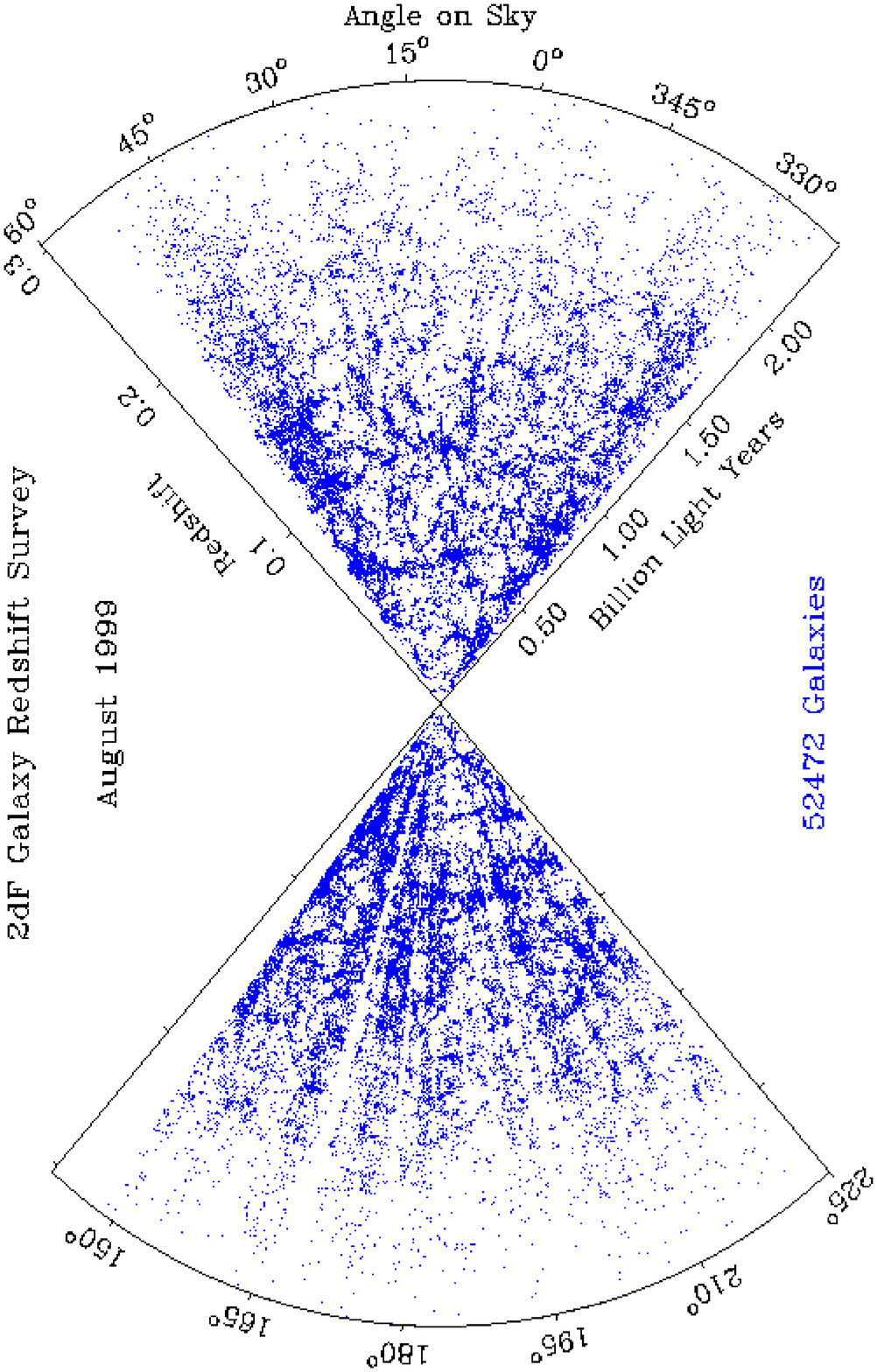,width=0.85\textwidth,angle=270}}
\vspace{15pt} \\
\parbox{0.85\textwidth}{\psfig{file=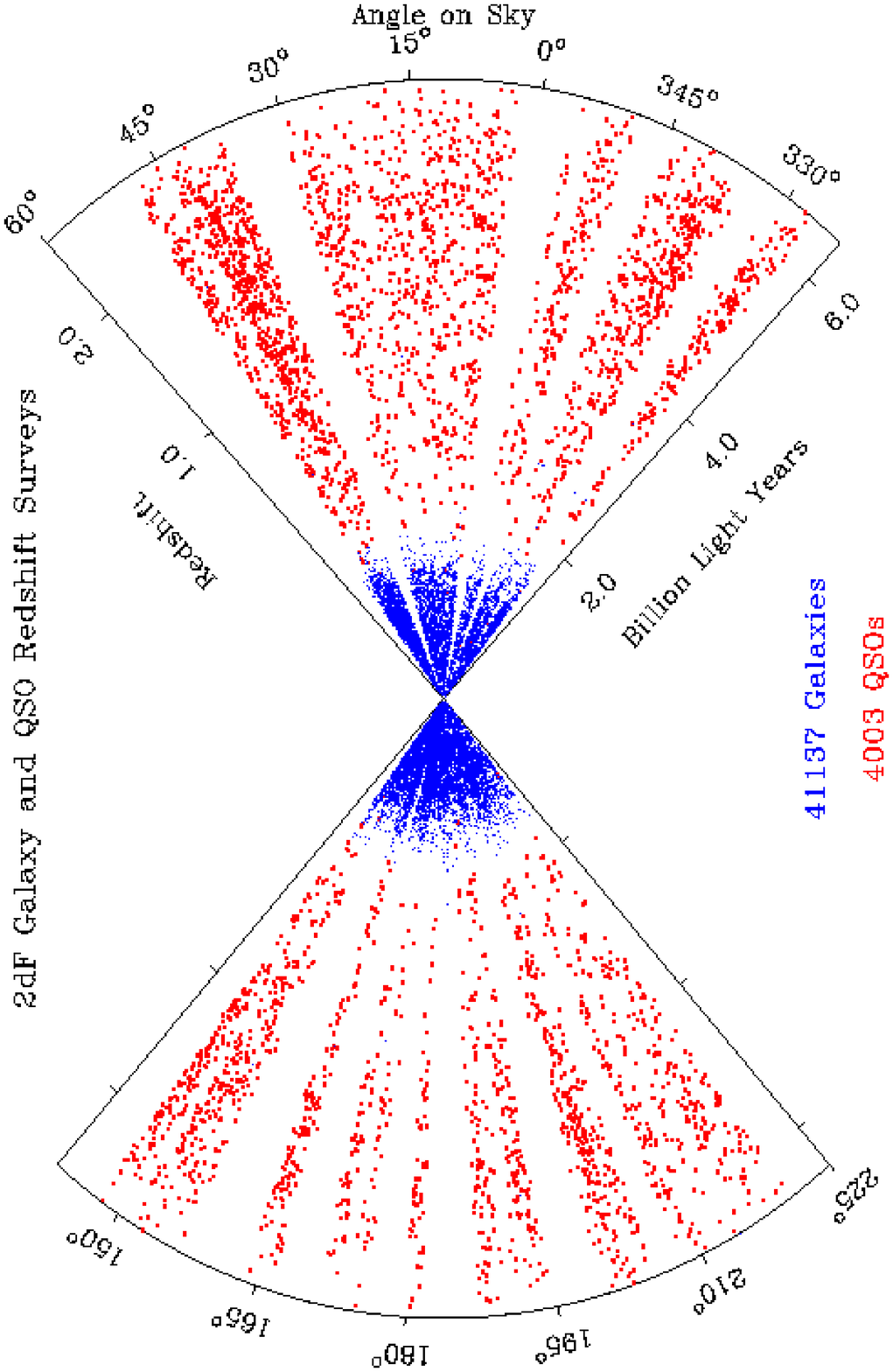,width=0.85\textwidth,angle=270}}
\vspace{15pt} \\
\caption{Redshift cone diagram showing the distribution of objects from
the 2dF galaxy and QSO redshift surveys as of August 1999. The top panel
shows the distribution of the galaxies alone, and the bottom panel shows
the distributions of both galaxies and QSOs.}
\label{fig:2dFzcone}
\end{center}
\end{figure}

The 2dFGRS has a companion survey in the \htmlink{2dF QSO Redshift
Survey}{http://www.mso.anu.edu.au/~rsmith/QSO_Survey} described by
\proclink{Smith}, which aims to obtain redshifts for more than 25,000
QSOs with 18.25$<$$b_J$$<$20.85 in two strips covering 740~sq.deg (see
Figure~\ref{fig:onsky}). The primary goals of the survey (\adslink{Boyle
et~al.\ 1999}{1999ldss.work...16}) are to measure the primordial
fluctuation power spectrum on very large scales, to obtain a geometric
determination of the cosmological constant $\Lambda$, and to investigate
the evolution of the QSO luminosity function and QSO clustering.
Follow-ups will include studies of absorption-line and damped
Lyman-$\alpha$ systems along QSO sight-lines. The multicolour selection
criteria for the target sample are expected to give a 93\% completeness
for the redshift range 0.3$<$$z$$<$2.2. As of August 1999, 13,854 of the
48,078 QSO candidates have been observed, and have yielded 7026
identified QSOs. Figure~\ref{fig:2dFzcone} shows both the 2dF galaxy and
QSO surveys on a single redshift cone diagram; the density of sampling
of the galaxies at low redshift is complemented by the sparser QSO
sample reaching out to beyond $z$=2.2.

The Sloan Digital Sky Survey (\htmlink{SDSS}{http://www.sdss.org}) will
be one of the largest and most comprehensive astronomical datasets ever
produced. Overviews of SDSS science and operations were given by
\proclink{Margon} and \proclink{Szalay} (see also \astroph{Margon
1999}{9805314}). SDSS is both a photometric and a spectroscopic survey
of $\pi$ steradians of the northern sky with $b>|30\degree|$. The survey
uses a purpose-built 2.5m telescope at Apache Point Observatory in New
Mexico, supported by a 0.5m telescope providing continuous monitoring of
the photometric conditions. The imaging system consists of a
high-efficiency drift-scan camera using 30 CCDs to image in five
passbands quasi-simultaneously while covering the sky in 3\degree-wide
strips with an effective exposure time of 55s. This will yield images
for 10$^8$ galaxies and 5$\times$10$^7$ stars down to $r^\prime$=23.1
over a quarter of the sky. The unique SDSS photometric system
($u^\prime$$g^\prime$$r^\prime$$i^\prime$$z^\prime$) was detailed by
\proclink{Fukugita}. The filter set is specially designed, featuring
high-throughput, nearly-disjoint passbands covering the range from
0.3--1.0$\mu$m. The imaging survey, which achieved first light in May
1998, has already produced exciting discoveries, including the first
$z$=5 QSO (\adslink{Fan et al.\ 1999}{1999AJ....118....1F}) and the
coolest methane brown dwarf (\adslink{Strauss et al.\
1999}{1999ApJ...522L..61S}). \proclink{Margon} listed the scientific
strengths of SDSS, in ascending order of importance, as: (i)~the large
database; (ii)~the homogeneity of the data; (iii)~the well-characterized
sample; (iv)~discovery potential; (v)~the archival value of the imaging
archive.

The volume of data to be generated by SDSS is enormous: 40~Terabytes of
raw data, reduced to a `mere' 1~Terabyte when processed.
\proclink{Szalay} described the complex data processing pipelines and
archival database which will generate and make available the SDSS data
products, ranging from a heavily compressed all-sky map, through
thumbnail images and spectra for every object, to the final photometric
and redshift catalogues. Szalay also outlined the sophisticated indexing
procedures which will be used to facilitate a rapid response to complex
search queries.

\proclink{Loveday} described the SDSS spectroscopic survey that will be
carried out in parallel with the imaging survey (\adslink{Loveday \&
Pier 1998}{1998wfsc.conf..317L}). There are various spectroscopic target
samples: the main galaxy sample, of 900,000 galaxies with Petrosian
magnitudes brighter than $r^\prime$$\approx$18.1; the Bright Red Galaxy
sample, of 100,000 galaxies brighter than $r^\prime$$\approx$19.3, with
selection by colour and photometric redshift to give early-type galaxies
with 0.25$<$$z$$<$0.45; the QSO sample, of 100,000 QSO candidates down
to $r^\prime$$\approx$20 selected by colour or as FIRST or ROSAT
sources; and the stellar sample, 100,000 stars of various types. In
addition there will be a smaller number of serendipity sources, selected
as `interesting' by a variety of criteria. It is important to note that
the redshift sample generated by this spectroscopic survey will be
supplemented by photometric redshifts with a precision of $\Delta
z$$\approx$0.05 for perhaps 10$^8$ fainter galaxies, which, in
conjunction with `spectral' types based on precision colours, will be
invaluable for studying the evolution of the galaxy population out to
$z$$\sim$1. A small amount of test data has been taken with the SDSS
spectrograph, a 640-fibre manually-configured system with a 3\degree\
field of view. Survey observations are expected to begin in early 2000,
with the whole SDSS, both photometric and spectroscopic surveys, taking
5~years to complete.

Another view of the nearby universe is provided by the Two-Micron
All-Sky Survey (\htmlink{2MASS}{http://www.ipac.caltech.edu/2mass}),
whose current status was reviewed by \proclink{Huchra}. 2MASS is
surveying the whole sky in the near-infrared $J$, $H$ and $K_s$ bands,
and for galaxies expects to achieve limiting magnitudes in these bands
of 15.0, 14.3 and 13.5 respectively (\adslink{Chester et~al.\
1998}{1998AAS...192.5511C}). The northern 2MASS survey began in 1997, and
the southern survey in 1998. At present 70\% of the sky has been covered
by 2MASS, and the whole survey is expected to be completed by early
2001.

The 2MASS imaging survey will be followed up by two (coordinated)
redshift surveys and two (complementary) peculiar velocity surveys.
\proclink{Huchra} described the \htmlink{2MASS Redshift
Survey}{http://cfa-www.harvard.edu/~huchra/2mass}, which has the goal of
measuring redshifts for 125,000 galaxies covering 90\% of the sky down
to $K_s$=12.2. This would build on existing redshift catalogues and, in
the north, fill in the missing redshifts with the 1.3m at FLWO used for
the CfA redshift surveys. \proclink{Colless} reported plans for a
southern counterpart to this survey, using the 6dF fibre spectrograph
currently under construction for the AAO Schmidt telescope. The
\htmlink{6dF Galaxy Survey}{http://www.mso.anu.edu.au/~colless/6dF}
(\astroph{Mamon 1998}{9908163}) has two phases, the first of
which is a redshift survey of 115,000 galaxies down to $K_s$=13,
$H$=13.5 or $J$=14.3 over the 17,000~sq.deg of the southern sky with
$b>|10\degree|$. As well as completing the all-sky 2MASS survey to
$K_s$=12.2 and pushing another 0.8~mag fainter, this first phase of the
6dF survey will provide a volume-limited sample for the second phase, a
peculiar velocity survey of at least 15,000 early-type galaxies at
distances less than $cz$=15000\kms\ using the Fundamental Plane distance
estimator. This in turn will complement an all-sky peculiar velocity
survey of 10,000 spirals using the infrared Tully-Fisher relation
proposed by \proclink{Huchra}.

\proclink{Webster} reviewed the progress and some preliminary results
from the HI Parkes All-Sky Survey
(\htmlink{HIPASS}{http://www.atnf.csiro.au/research/multibeam/.research.html};
\adslink{Staveley-Smith et~al.\ 1999}{1999ldss.work..132S}), which is
currently using the Parkes 21cm multi-beam receiver to cover the sky
south of $\delta$=+25\degree\ out to a maximum redshift of
$cz$=12700\kms. As of August 1999, the southern hemisphere survey is
over 90\% complete, and the northern extension is over 10\% complete.
HIPASS provides a complete 21cm survey of the southern sky with 7~arcmin
angular resolution down to a 3$\sigma$ HI mass limit of
1.4$\times$$10^{10}(D/100\Mpc)^2M_\odot$ (\adslink{Kilborn et~al.\
1999}{1999PASA...16....8K}). When complete, HIPASS is expected to contain
$\sim$5000 galaxies with a mean redshift of 3000\kms. The effective
volume of the sample is 6$\times$10$^6$\cubicMpc. HIPASS should offer a
very different view of the local universe to optical and near-infrared
redshift surveys, and it will be interesting to see the differences in
the galaxy population and the large-scale structure, especially for
low-mass objects.

\begin{figure}
\begin{center}
\parbox{0.55\textwidth}{\psfig{file=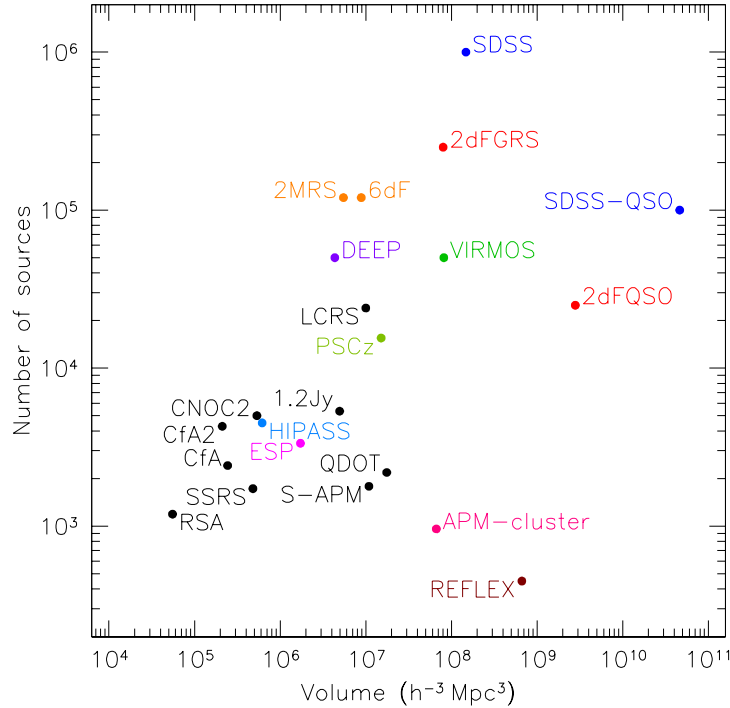,width=0.55\textwidth}}
\caption{Comparison of the size and volume of existing and planned
redshift surveys. Surveys are labelled by their standard acronyms (see
text).}
\label{fig:sizevol}
\end{center}
\end{figure}

Figure~\ref{fig:sizevol} compares the number of objects and the volume
encompassed by various redshift surveys.

\section{Surveys at High Redshift}

The Deep Extragalactic Evolutionary Probe
(\htmlink{DEEP}{http://www.ucolick.org/~deep}) is a two-phase survey of
galaxies at redshifts out to $z$$\sim$1 using the Keck Telescopes and
HST. \proclink{Koo} reported results from the first phase of DEEP, now
approaching completion, and outlined plans for the more ambitious second
phase, which will use the
\htmlink{DEIMOS}{http://www.ucolick.org/~loen/Deimos/deimos.html}
spectrograph on Keck. The first phase (\adslink{Koo
1998}{1998wfsc.conf..161K}) utilised LRIS on Keck, in conjunction with
HST imaging, to carry out a redshift survey of $\sim$1000 galaxies down
to $I$=24. This survey was followed up with more detailed studies of
galaxy kinematics and other physical properties for small subsamples.
The second phase, using the multi-slit spectrograph DEIMOS, aims to
perform a redshift survey of 50,000 galaxies to $I$=23 plus another 5000
galaxies going one magnitude deeper (\adslink{Davis \& Faber
1998}{1998wfsc.conf..333D}). The sample will be defined using
photometric redshifts to select the objects with 0.7$<$$z$$<$1.2. The
main goals of the survey will be to study the evolution of both the
galaxy population and the large-scale structure, and it is expected to
run from early 2000 to late 2003.

More ambitious still is the \htmlink{VIRMOS}
{http://www.astrsp-mrs.fr/www_root/projets/virmos/virmos-top.html}
survey described by \proclink{Guzzo}. This will use the optical (VIMOS)
and near-infrared (NIRMOS) multi-slit spectrographs currently being
built for ESO's VLT. Using 120 nights of guaranteed time on the VLT, the
survey will measure redshifts for 100,000 galaxies down to $I_{AB}$=22
over 20~sq.deg, and for 50,000 galaxies down to $I_{AB}$=24 over
2~sq.deg (\adslink{LeFevre et~al.\ 1999}{1999obco.conf..250L}). A
further $\sim$1000 redshifts will be obtained down to $I_{AB}$$\sim$26
over a 1~sq.arcmin field using the spectrograph's 6400-fibre integral
field unit. VIMOS will see first light in mid-2000 and NIRMOS in
mid-2001; the survey is expected to take 3--4~years.

\section{Large-Scale Structure}

Results on large-scale structure (LSS) were reported by a number of
speakers at the conference. \proclink{Sutherland} described a
determination of the power spectrum of galaxy fluctuations from the
IRAS-based Point Source Catalogue redshift survey
(\htmlink{PSCz}{http://www-astro.physics.ox.ac.uk/~wjs/pscz.html}).
Using a sample of 14,500 galaxies with 60$\mu$m flux of more than 0.6~Jy
covering 84\% of the sky, the PSCz team find that they are able to
determine the power spectrum, $P(k)$, for wavenumbers down to
0.03\invMpc (i.e.\ scales up to 200\Mpc). The results
(\adslink{Sutherland et~al.\ 1999}{1999MNRAS.308..289S}, \adslink{Tadros
et~al.\ 1999}{1999MNRAS.305..527T}) are consistent with the earlier QDOT
and 1.2~Jy surveys, although $P(k)$ is significantly better determined.
For plausible values of the small-scale velocity dispersion, $P(k)$ is
well-fit by a CDM-like model with shape parameter $\Gamma$$\approx$0.25
and normalisation $\sigma_8$$\approx$0.75 (although assuming a larger
dispersion permits a model with $\Gamma$$\approx$0.5 to be marginally
acceptable).

\proclink{Rowan-Robinson} reported an estimate of $\beta=\Omega^{0.6}/b$
derived from the comparison of the Local Group (LG) motion with respect
to the CMB and the motion predicted from a model based on clusters and
voids identified in the PSCz and the literature (see also
\adslink{Schmoldt et~al.\ 1999a}{1999MNRAS.304..893S},
\adslink{1999b}{1999AJ....118.1146S}). The predicted LG dipole has
nearly converged by $z$=0.1 (with most of the LG motion generated within
200\Mpc); linear theory requires $\beta$$\approx$0.7 to fit the observed
amplitude of the LG motion.

\proclink{Guzzo} summarised the results of the ESO Slice Project
(\htmlink{ESP}{http://boas5.bo.astro.it/~cappi/esokp.html};
\astroph{Vettolani et~al.\ 1997}{9704097}) redshift survey. This
pre-cursor to surveys like the 2dFGRS used photographic sky survey
plates to select a sample of galaxies down to $b_J$=19.4 in a single
1\degree$\times$35\degree\ strip. Redshifts were obtained for 3342
galaxies (85\% complete; Vettolani et al.\ 1998). Perhaps the most
significant LSS result to emerge from ESP, is that there is good
evidence for a local under-density of nearly a factor of 2 extending out
to 250\Mpc\ (\astroph{Zucca et~al.\ 1997}{9705096}). This is consistent
with the normalisation difference between the luminosity functions
derived from surveys at this depth (ESP itself and the Autofib survey,
\adslink{Ellis et~al.\ 1996}{1996MNRAS.280..235E}) and those derived
from shallower surveys such as APM-Stromlo (\adslink{Loveday et~al.\
1992}{1992ApJ...390..338L}). It also `explains' the origin of the steep
number counts at bright magnitudes (\adslink{Maddox et~al.\
1990}{1990MNRAS.247P...1M}), although it begs the question of the
completeness of the galaxy catalogues derived from photographic surveys
at bright magnitudes. The local void implied by this result would appear
to cover much of the south Galactic cap, as evidenced by the north/south
difference in galaxy density found in the PSCz, LCRS and CfA2 surveys.
Is this void, with an amplitude $\delta\rho/\rho$$\sim$$\approx$0.5
compatible with statistical measures of clustering? The ESP
redshift-space correlation function is in good agreement with other
determinations on scales between 1 and 50\Mpc, though a little lower on
the smaller scales. The power spectrum is likewise consistent with other
determinations at wavenumbers $k$$>$0.1\invMpc. However the small volume
covered by the survey, and its strip geometry, prevent a reliable
determination of $P(k)$ at wavenumbers below about 0.1\invMpc\ (i.e.\
scales above 60\Mpc).

An efficient approach to surveying larger volumes, and so measuring
$P(k)$ at smaller $k$, is to use redshift surveys of clusters, since
clusters are about 5$\times$ more strongly clustered than galaxies.
Results from two such surveys were summarised at the meeting.
\proclink{Boehringer} and \proclink{Guzzo} reported early results from
the REFLEX survey, while \proclink{Tadros} summarised the results from
the first stage of the APM cluster survey.

\proclink{Boehringer} noted the advantages of an X-ray selected
cluster sample as being the close correlation between $L_X$ and mass,
and the minimisation of projection effects. He described the detection
of clusters in the ROSAT All Sky Survey for the REFLEX (southern) and
NORAS (northern) samples (\adslink{Guzzo et~al.\
1999}{1999Msngr..95...27G}). Together these samples comprise over 900
clusters covering the whole sky at latitudes $b>|20\degree|$ down to a
flux limit of $F_X$=3$\times$10$^{-12}$~erg\,s\,cm$^2$. The preliminary
power spectrum derived from the REFLEX survey using a subset of 188
clusters in a 400\Mpc\ co-moving cube (\adslink{Schuecker et~al.\
1998}{1998tx19.confE.546S}) appears to show a significant turnover in
$P(k)$ at $k$$\approx$0.04--0.05\invMpc, corresponding to 125--160\Mpc.
\proclink{Guzzo} presented very preliminary results from a larger sample
of clusters in a 1000\Mpc\ comoving cube which confirm the turnover in
$P(k)$. But the newer results also hint at a possible second peak in
$P(k)$ at $k$$\sim$0.01, although a $\Omega_\Lambda$=0.7 flat CDM model
remains consistent within the (large) uncertainties.

\proclink{Tadros} described a similar survey, the APM Cluster Redshift
Survey (\adslink{Tadros et~al.\ 1998}{1998MNRAS.296..995T}). The first
phase of this survey consisted of 364 clusters drawn from the APM
cluster catalogue of 960 clusters covering 4500~sq.deg. The APM clusters
provide a cleanly-selected sample that is largely unaffected by the
inhomogeneities and projection effects afflicting the Abell catalogue.
The redshift-space cluster-cluster correlation function $\xi_{cc}(s)$ is
well represented by a power law with index $\gamma$$\sim$2 and a
correlation length of 14\Mpc\ (smaller than that of the Abell
catalogue). The LSS as measured from clusters provides a clean
comparison with cosmological models, since clusters are readily
identified from the dark matter halos without the confusion of a bias
parameter. The APM $\xi_{cc}(s)$ is consistent with $\Lambda$CDM and MDM
models, but has more clustering than is predicted by standard CDM.

An estimate of $\beta=\Omega^{0.6}/b$ can be obtained by comparing the
real-space and redshift-space cluster-galaxy cross-correlation functions
$\xi_{cg}$. The real-space $\xi_{cg}(r)$ was derived from the cluster
survey and the APM galaxy catalogue by inverting the angular
cluster-galaxy cross-correlation function. The redshift-space
$\xi_{cg}(\sigma,\pi)$, as a function of separation in the plane of the
sky ($\sigma$) and along the line of sight ($\pi$), was derived from the
cluster survery and the Stromlo-APM redshift survey. These two
$\xi_{cg}$'s can be related using a model incorporating non-linear
infall and the velocity dispersion of galaxies around clusters.
\proclink{Tadros} showed that the best-fit model yields an estimate of
$\beta$$\approx$0.4, with a 95\% upper confidence limit of
$\beta$$<$0.7. The power spectrum from the APM cluster survey shows a
turnover at $k$$\approx$0.03\invMpc, but the median depth of the survey
(270\Mpc) is only just adequate to measure $P(k)$ on this 200\Mpc\
scale. To improve the significance of this detection and reduce the
systematic errors, the second phase of the APM cluster survey, now
underway, will obtain redshifts for the remaining clusters in the APM
cluster catalogue, bringing the sample up to 960 clusters. This cluster
survey, which will take 2--3 years to complete, will be particularly
useful since it overlaps with the 2dF Galaxy Redshift Survey.

The {\em faint} galaxy correlation function at large angular scales was
discussed by \proclink{Brown}, who has used digitally-stacked Schmidt
plates to achieve an approximate limit of $B_J$=23.5. This gives 700,000
galaxies in each of two 40~sq.deg fields. The median redshift at this
depth is $z$$\approx$0.4. The correlation function he derives is
well-fit as a power-law, $\omega(\theta)\propto\theta^{1-\gamma}$ with
$\gamma=1.7$, over the range 0.05--10\Mpc. The amplitude of the
correlation function declines as $(1+z)^{-(3+\epsilon)}$ with
$\epsilon$$\approx$0, corresponding to fixed clustering in physical
coordinates. However these results hide the different clustering
properties of the red and blue galaxies, which are respectively fitted
by models with $\gamma$=1.8, $\epsilon$=$-$1.3 and scalelength
$r_0$=8.6\Mpc, and $\gamma$=1.6, $\epsilon$=$-$1.5 and $r_0$=3.5\Mpc.
The clustering amplitude of red galaxies is thus about 5$\times$ higher
than the blue galaxies. The lack of any significant strengthening of the
clustering amplitude of blue galaxies out to $z$$\sim$0.4 suggests that
the increase in the population of blue galaxies in clusters with
redshift (the Butcher-Oemler effect) is simply related to the overall
increase in the faint blue galaxy population with redshift.

The LSS goals of the
\htmlink{2dFGRS}{http://www.mso.anu.edu.au/~colless/2dF}, and some
preliminary results from the survey, were described by
\proclink{Dalton}. The main LSS goals are: (i)~the determination of
$P(k)$ on large scales ($>$100\Mpc); (ii)~the topology of the 3D
distribution; (iii)~tests of gaussianity (on large scales) and biasing
(on small scales) from higher-order statistics; (iv)~estimation of the
mass density $\Omega$ and the bias parameter(s) $b$ from the
redshift-space distortions. Dalton discussed the tiling and
fibre-assignment algorithms used in the survey, and showed that the
restrictions on minimum fibre separations mean the survey is
significantly biased against close pairs with separations less than
2~arcmin (corresponding to 150\kpc\ at the median depth of the survey,
$z$=0.1). This effect, and the variation in redshift completeness with
apparent magnitude, are taken into account when estimating or simulating
the large-scale structure statistics derived from the survey.

\begin{figure}
\begin{center}
\parbox{0.51\textwidth}{\psfig{file=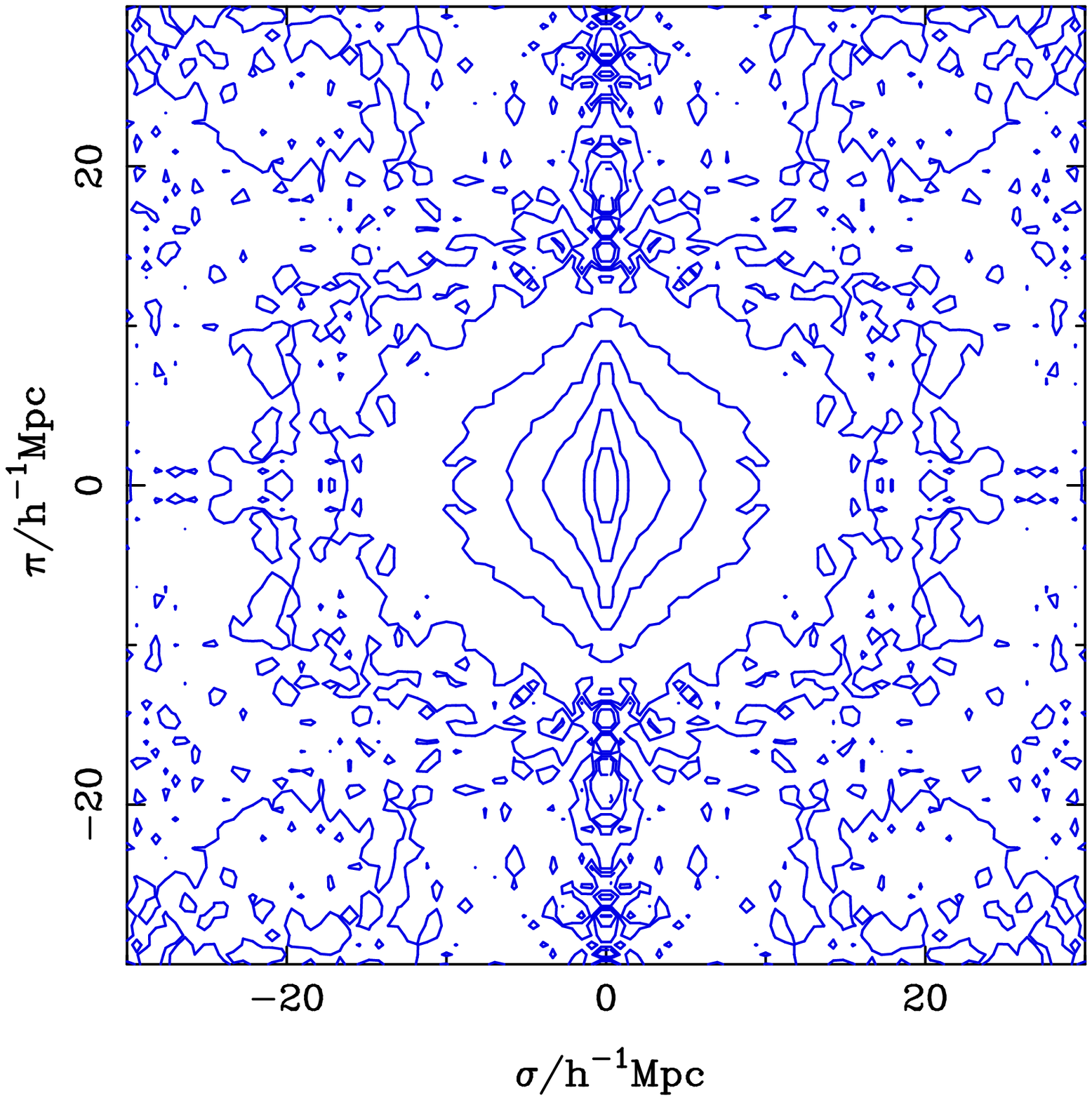,width=0.51\textwidth}}
\hfill
\parbox{0.44\textwidth}{\psfig{file=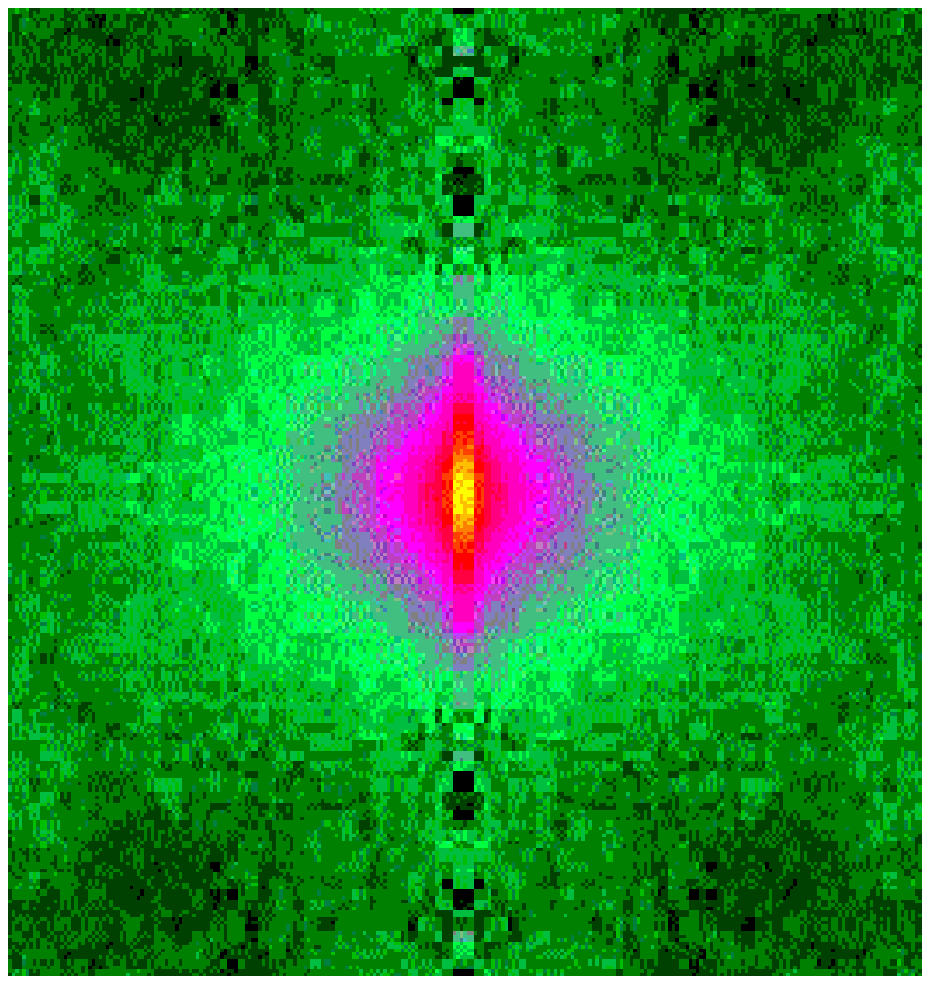,width=0.44\textwidth}}
\caption{The 2dFGRS redshift-space correlation function,
$\xi(\sigma,\pi)$, shown both as a contour plot (left) and false-colour
image (right). Note the stretching/flattening along the line of sight
($\pi$) for small/large separations in the plane of the sky ($\sigma$).}
\label{fig:xisigpi}
\end{center}
\end{figure}

\proclink{Dalton} showed preliminary determinations of the
redshift-space correlation function $\xi(\sigma,\pi)$ for the 2dFGRS.
There is very good agreement with the results obtained from the Las
Campanas Redshift Survey
(\htmlink{LCRS}{http://manaslu.astro.utoronto.ca/~lin/lcrs.html}). The
distortions in $\xi(\sigma,\pi)$ shown in Figure~\ref{fig:xisigpi} are
roughly consistent with a model with $\beta=\Omega^{0.6}/b\approx0.5$
and small-scale velocity dispersion $\sigma\approx400\kms$. A similar
value for $\beta$ emerges from the quadrupole to monopole ratio of the
redshift-space distortions on large scales. As a consistency check on
the results, the projected correlation function $\Xi(r)$, derived by
integrating over $\xi(\sigma,\pi)$, has been compared to that obtained
by \adslink{Baugh \& Efstathiou (1993)}{1993MNRAS.265..145B} from a
deprojection of the angular correlation function of the parent APM
galaxy catalogue. There is excellent agreement on scales up to 30\Mpc,
while on larger scales the cosmic variance dominates the uncertainties
in the as-yet-incomplete 2dFGRS.

\begin{figure}
\begin{center}
\parbox{0.48\textwidth}{\psfig{file=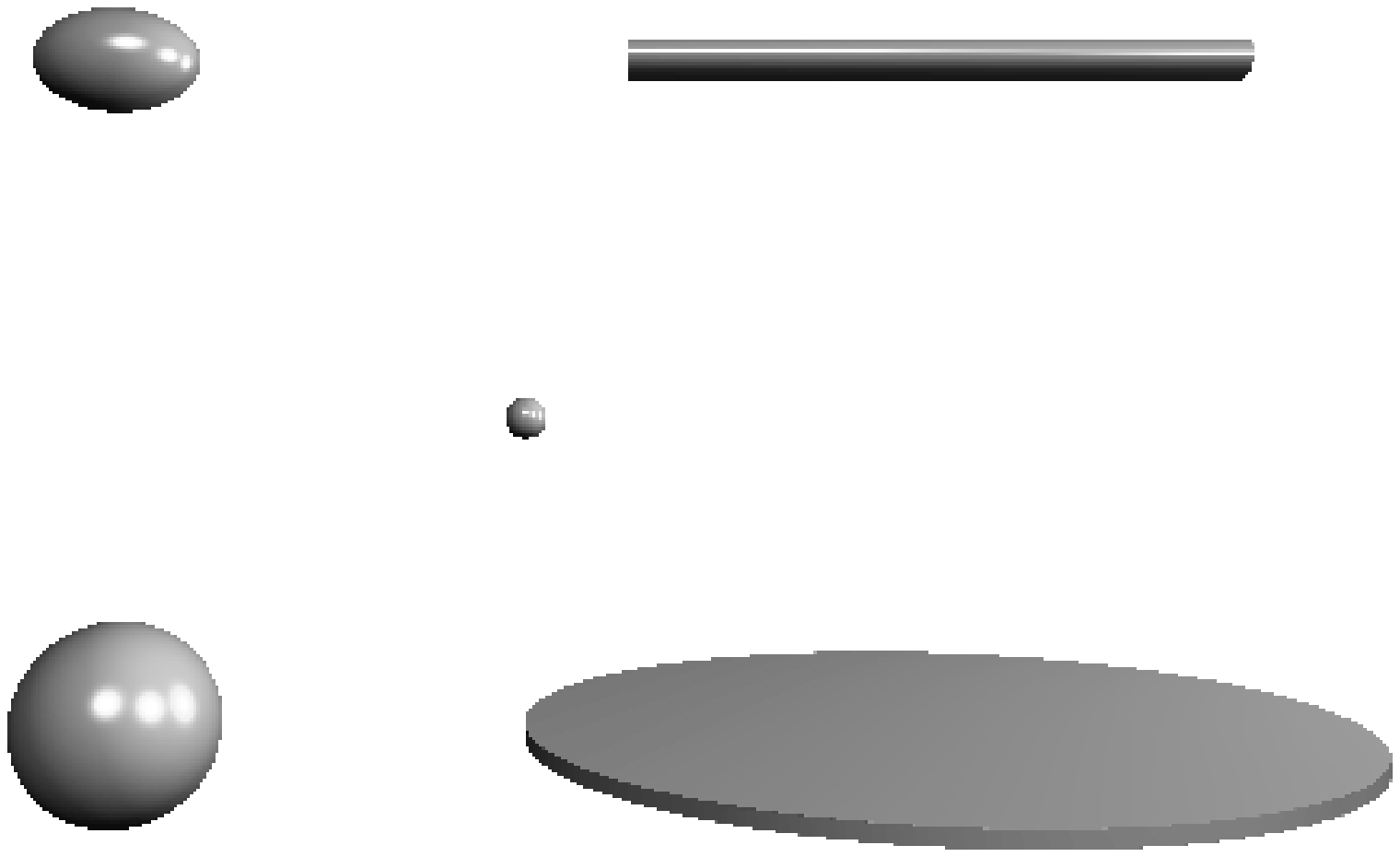,width=0.48\textwidth}}
\hfill
\parbox{0.48\textwidth}{\psfig{file=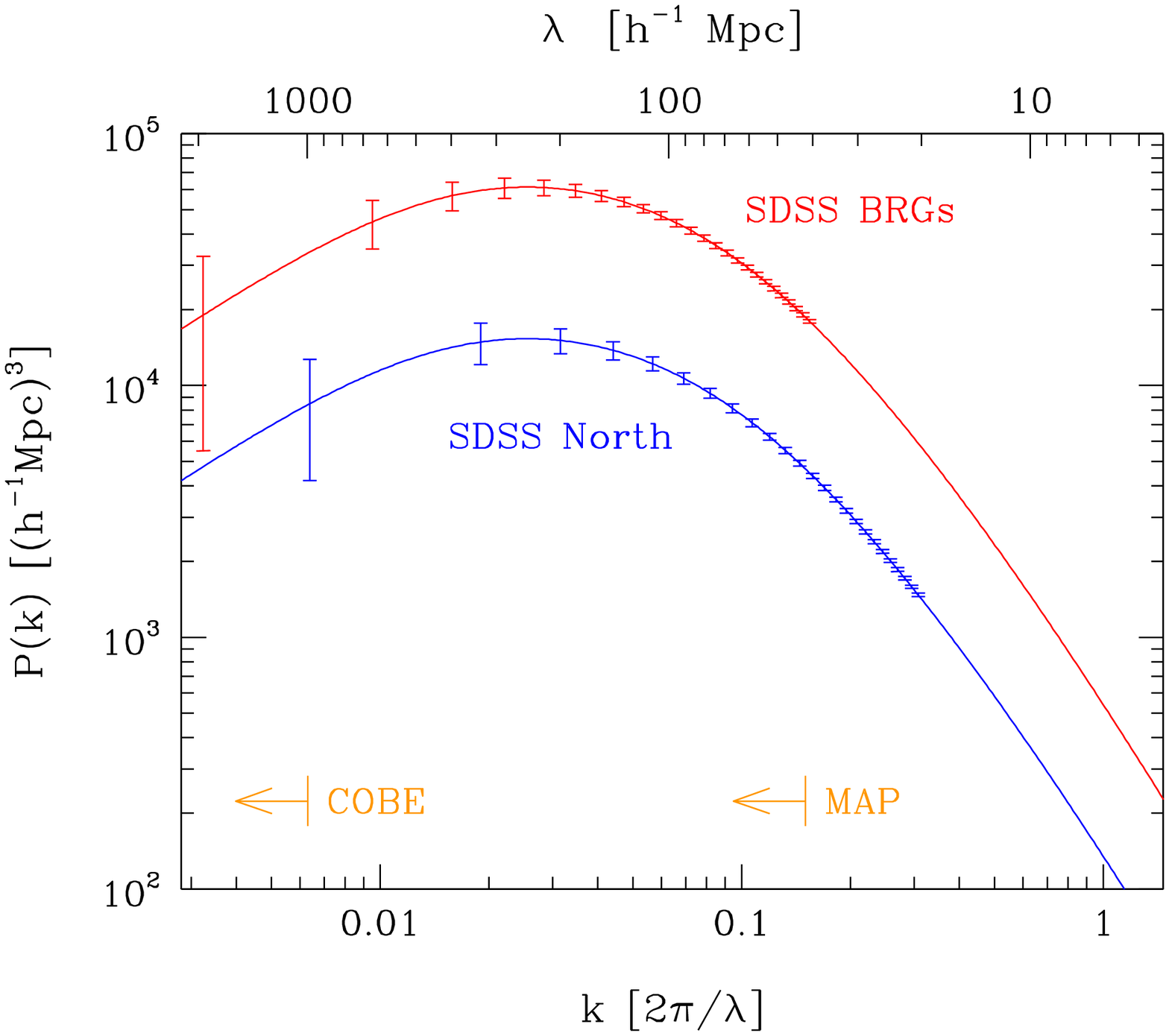,width=0.48\textwidth}}
\caption{(a)~The window function functions for various surveys: lower
left is QDOT, upper left is CfA2, upper right is LCRS, lower right is
the BEKS pencil-beam survey, and the small dot in the center is SDSS and
2dFGRS (from the SDSS \htmlink{Black
Book}{http://www.astro.princeton.edu/BBOOK}). (b)~The prediction for the
recovered power spectrum and estimated errors from the SDSS main
northern galaxy sample and from the Bright Red Galaxy sample
(\adslink{Loveday \& Pier 1998}{1998wfsc.conf..317L}). The scales
covered by the COBE and MAP cosmic microwave background probes are also
indicated.}
\label{fig:wkpk}
\end{center}
\end{figure}

In a similar vein, \proclink{Szalay} showed the results of simulations
indicating the precision that the Sloan survey will achieve in
recovering cosmological parameters from LSS statistics. SDSS should be
able to measure the power spectrum normalization $\sigma_8$ to a
precision of 5\%, the power spectrum shape parameter $\Gamma$ to 20\%,
and the redshift-distortion parameter $\beta$ (which involves both the
mass density $\Omega$ and the bias parameter $b$,
$\beta=\Omega^{0.6}/b$) to 35\%. The precision with which $P(k)$ will be
determined is illustrated in Figure~\ref{fig:wkpk}, which shows both the
effective window functions of various surveys and the simulated recovery
of $P(k)$ from the Sloan survey, with estimated errors.

On larger scales of both space and time, \proclink{Croom} reported
preliminary LSS results for the \htmlink{2dF QSO Redshift
Survey}{http://www.mso.anu.edu.au/~rsmith/QSO_Survey}. The LSS goals of
the QSO survey are: (i)~determining the QSO $P(k)$ out to scales
$\sim$1000\Mpc; (ii)~measuring the cosmological constant $\Lambda$ from
geometrical (as opposed to dynamical) distortions of clustering in
redshift space; (iii)~tracing the evolution of QSO clustering out to
$z$$\sim$3 to constrain $\Omega$ and the QSO bias parameter. As with the
galaxy survey, corrections are needed for the partial coverage of
overlapping fields, the deficit of close pairs and Galactic extinction.
Preliminary determinations of the 0.3$<$$z$$<$2.2 QSO correlation
function using 2765 QSOs give $\gamma$$\approx$1.4 and scaling lengths
$r_0$$\approx$3\Mpc\ and for $r_0$$\approx$5\Mpc\ for models with
($\Omega_M$=1,$\Omega_\Lambda$=0) and
($\Omega_M$=0.3,$\Omega_\Lambda$=0.7) respectively, with the clustering
appearing to be constant in comoving coordinates over this range. With
the full QSO survey it should be possible to measure $\gamma$ to 3\% and
$r_0$ to 5\% on small scales, and on large scales to measure $P(k)$ on
scales $k$$>$0.01\invMpc\ (i.e.\ $<$600\Mpc).

Other cosmological parameters were addressed by Mould and Peterson.
\proclink{Mould} summarised the results from the HST Key Project to
measure the Hubble constant, $H_0$ (\astroph{Mould et~al.\
1999}{9909260}). This work essentially consists of using Cepheid
distances to calibrate a wide variety of distance estimators, including
type Ia supernovae, the Tully-Fisher relation for spiral galaxies, and
the Fundamental Plane and surface brightness fluctuations for
bulge-dominated galaxies. The available data from all these estimators
yield consistent values for $H_0$, and the combined best estimate, after
correcting for the chemical composition dependence of the Cepheid
period-luminosity relation, is $H_0 = (68 \pm
6)$\,km\,s$^{-1}$\,Mpc$^{-1}$ (including random and systematic errors),
for an assumed LMC distance of 50$\pm$3\,kpc. \proclink{Rowan-Robinson}
also gave an estimate of the Hubble constant using a similar compilation
of methods based on the Key Project Cepheid distances, but with the
addition of corrections for peculiar velocities based on the PSCz flow
model. He finds $H_0 = (65 \pm 2)$\,km\,s$^{-1}$\,Mpc$^{-1}$ (random
error only).

\proclink{Mould} also considered the possibility that we inhabit a large
local void with $-0.5<\delta n/n<-0.2$, as suggested by \proclink{Guzzo}
from the results of the ESP survey, and as is consistent with the
preliminary results from the 2dFGRS. Since $\delta H_0/H_0 =
\frac{1}{3}\beta\delta n/n$ this would imply (for $\beta =
\Omega^{0.6}/b \approx 0.5$) that $0.92<H_0^{\rm global}/H_0^{\rm
local}<0.97$.

\proclink{Peterson} addressed the question of whether a cosmological
constant is demanded by the galaxy number counts. He finds that the
surface density of faint galaxies derived from the optical and
near-infrared number counts is too high to be compatible with a
$\Omega_M$=1 cosmology, and are much better fitted by a low-density flat
universe with $\Omega_\Lambda$$\approx$0.8. Although this claim depends
on the assumed evolutionary history of the galaxies, Peterson argued
that models which reproduce the number counts by invoking merger-driven
evolution are inconsistent with the low measured amplitude of the
angular correlation function for faint galaxies.

\section{Galaxy Population and Evolution}

The new generation of large redshift surveys promise to reveal the
properties of the galaxy population in extraordinary detail. Some of
this potential was revealed by various preliminary results emerging from
the 2dFGRS. \proclink{Colless} summarised some of the recent work by
members of the survey team on the local luminosity function (LF). The
overall LF shows clear signs of being more complex than the standard
Schechter form, with a steepening of the faint end for $M_{b_J}>-16$.
Using a principal component analysis (PCA) to classify the galaxies into
5 spectral types (\adslink{Folkes et~al.\ 1999}{1999MNRAS.308..459F}),
it is clear that this is the result of a general steepening of the faint
end for later-type galaxies, which dominate the population at lower
luminosities.

However this type of analysis does not help explain the large variation
between different determinations of the local LF, not only in the faint
end but also in the normalization. A more general approach is required,
which will account for surface brightness selection effects.
\proclink{Colless} reported a preliminary determination (Cross et~al.,
in prep.) of the bivariate brightness distribution (BBD), the joint
distribution over absolute magnitude and surface brightness. This shows
that the 2dFGRS is {\em not} severely affected by its selection limit at
low surface brightness, as the galaxy number distribution falls off
sharply before that limit is reached. The BBD is strongly suggestive of
the existence of separate dwarf and giant populations, the former with a
LF that is flat or declining at the faint end and the latter with a
steeply increasing faint end.

More detailed analyses of the galaxy population can be carried out using
various quantities derived from the spectra: \proclink{Colless} showed
that measurement of the H$\alpha$ equivalent width can be used to map
the relative distributions of the star-forming and quiescent galaxies in
the 2dFGRS (Lewis et~al., in prep.), while \proclink{Deeley} used the
Balmer-line and [OII] equivalent widths to identify post-starburst E+A
galaxies, finding an incidence of 0.25\%.

A very powerful application of the redshift surveys is in the
identification of sources detected in sky surveys at other wavelengths.
\proclink{Cannon} reported on a program to identify the
\htmlink{NVSS}{http://www.cv.nrao.edu/~jcondon/nvss.html} and
\htmlink{SUMSS}{http://www.astrop.physics.usyd.edu.au/SUMSS} radio
galaxies in the 2dFGRS (\astroph{Sadler et~al.\ 1999}{9909171}). About
5\% of NVSS sources are found in the 2dFGRS, and about 1.5\% of 2dFGRS
sources are in NVSS; in the full redshift survey there should be about
4000 radio galaxies. The 700 sources identified so far, with roughly
equal numbers of star-forming galaxies, optical AGN and radio AGN,
already make up the largest homogeneous sample of optical spectra for
radio galaxies. The first result emerging from this sample is a
much-improved determination of the radio luminosity function, which has
a double-humped appearance. By using the PCA spectral types,
\proclink{Cannon} showed that this structure is due to different
luminosity functions for the AGN (which dominate at high luminosity) and
the star-forming galaxies (which have a steeper slope and dominate at
lower luminosities).

\proclink{Jackson} described the potentialities for future work
combining radio surveys such as NVSS, SUMSS and
\htmlink{FIRST}{http://sundog.stsci.edu/top.html} with large redshift
surveys such as the 2dF galaxy and QSO surveys, SDSS and the 6dF survey.
Understanding the properties of radio galaxy populations is very much
hindered by the lack of large, homogeneous redshift samples.
Cross-identification with the 2dFGRS allows the definition and detailed
studies of the local radio galaxy populations: the luminosity functions
(see above), environmental effects, clustering and low-redshift
evolution. Cross-identification with the 2dF QSO survey allows similar
studies of radio-loud QSOs. Limitations arise from the joint
radio/optical selection criteria, nonetheless such studies promise to
greatly improve our understanding of radio galaxies.

\proclink{Boyle} showed the results obtained on QSO evolution using a
homogeneous sample of 3265 QSOs from the 2dF QSO Redshift Survey. He
finds that the luminosity function is fitted by a double power law
(i.e.\ it has a break), and that the change in the LF with redshift is
consistent with a $\Omega_M$$\approx$0.3, $\Omega_\Lambda$$\approx$0.7
universe and pure luminosity evolution with the approximate form
$L^\ast(z)/L^\ast(0)\propto{\rm dex}(1.4z-0.27z^2)$ out to $z$=2.3. He
noted that the SDSS QSO survey, which is not limited by UVX selection,
may be able to see whether this evolution turns around for redshifts
between 2.3 and 3, with decreasing numbers of QSOs at fixed luminosity.
He also reported results from HST and ground-based imaging of a sample
of 76 low-redshift, X-ray selected AGN (Schade et~al.\ 1999, in prep.).
Host galaxies with $-23$$<$$M_{\rm host}$$<$$-17$ were detected in all
cases, with 55\% of AGN residing in bulge-dominated galaxies. A weak
correlation is found between the AGN and host luminosities.
Interestingly, 10\% of AGN hosts do not show a point source (in
agreement with HST studies of fainter AGN). The luminosities, sizes and
morphologies of the AGN hosts are completely consistent with the
`normal' galaxy population, except that AGN preferentially reside in
bulge-dominated systems.

\proclink{Giavalisco} presented some recent results from surveys of
Lyman-break galaxies (LBGs) at z$\sim$3 and z$\sim$4
(\adslink{Giavalisco et~al.\ 1998}{1998ApJ...503..543G},
\adslink{Steidel et~al.\ 1999}{1999ApJ...519....1S}). The
colour-selection of LBGs is a highly efficient method for generating
samples of star-forming galaxies at high redshift. Follow-up
spectroscopic observations have now provided redshifts for more than 750
galaxies with $z$=2--5. The selection biases are believed to be
well-understood, so that it is possible to compute the far-ultraviolet
luminosity function (LF) for the LBGs. Combining ground-based and Hubble
Deep Field samples, the $z$$\sim$3 LF (measured in the $R$ band, which
corresponds to $\sim$1700\AA\ in the rest-frame) is found to be
well-represented by a Schechter function with a steep faint end
($\alpha$$\approx$$-$1.6), although the slope does depend on the
incompleteness correction that is applied. With assumptions about the
LBG spectral energy distribution, and using the \adslink{Calzetti
(1997)}{1997AJ....113..162C} reddening law, it is possible to estimate
the extinction by dust in the LBGs. The median extinction at rest-frame
1500\AA\ is found to be 1.7~mag, and integrated over the whole
population the UV extinction is about a factor of 7. The smaller
spectroscopic sample of LBGs selected at $z$$\sim$4 does not show any
significant change from the $z$$\sim$3 population: the LFs of both
samples are consistent in shape and normalisation. The LBGs are very
strongly clustered, with co-moving correlation function amplitude at
least as great as that of $z$$\sim$0 galaxies. Given the weaker mass
clustering at early epochs, this implies much stronger biasing. However
this picture is consistent with the predictions of simple analytic
models (e.g.\ \adslink{Mo et~al.\ 1999}{1999MNRAS.304..175M}) for the
evolution of galaxy clustering based on the measured power spectrum for
low-redshift galaxies.

Francis and Illingworth reported studies of high-redshift galaxy
clusters. \proclink{Illingworth} showed HST results relating to the
nature of galaxies in rich environments at $z$$\sim$0.3--1. The mild
evolution in the Fundamental Plane for the E and S0 galaxies in the
cluster CL1358+62, relative to the nearby Coma cluster, implies that
these galaxies are a mature, homogeneous population whose stars formed
at redshifts $z$$>$1 (\astroph{Kelson et~al.\ 1999}{9906152}). The
absorption linestrengths are also consistent with old, single-burst
populations in which metallicity increases with velocity dispersion. For
early-type spirals, however, there is a positive correlation between
luminosity-weighted age and velocity dispersion. In contrast to this
evidence for a high redshift of star formation in early-type galaxies,
\proclink{Illingworth} also reported results from a morphological study
of the cluster MS1054$-$03 at $z$=0.83 which show a high merger rate in
this luminous X-ray cluster. If this rate is typical for clusters at
this redshift, it implies that up to 50\% of the massive ellipticals in
clusters may be formed in mergers at redshifts $z$$<$1 (\adslink{van
Dokkum et~al.\ 1999}{1999ApJ...520L..95V}). Hence, the stars in cluster
ellipticals form early, but the galaxies' final morphological forms may
be more lately acquired.

\proclink{Francis} described a detailed investigation of a candidate
galaxy cluster at $z$=2.38. Narrow-band imaging of this cluster shows
three bright Lyman-$\alpha$ sources, one of which is more than 50\kpc\
in extent. Two of these sources have very red colours, and near-infrared
photometry shows that their spectral energy distributions are poorly-fit
by dust-obscured star-forming regions, but well-fit by a stellar
populations with ages $\sim$0.5\,Gyr and masses of a few times
10$^{11}$\Msol . A number of other objects with similar infrared colours
are seen in the cluster, and have luminosity profiles more consistent
with an $R^{1/4}$ law than an exponential disk. The cluster lies in
front of three QSOs, and in all three sight-lines there are neutral
hydrogen absorption lines (including damped Lyman-$\alpha$ systems) at
the cluster redshift. This suggests the cluster contains a large amount
of neutral hydrogen, perhaps more than 10$^{12}$\Msol, which may be in
the form of large numbers of short-lived 10$^6$\Msol\ clouds.

\begin{figure}
\begin{center}
\parbox{0.47\textwidth}{\psfig{file=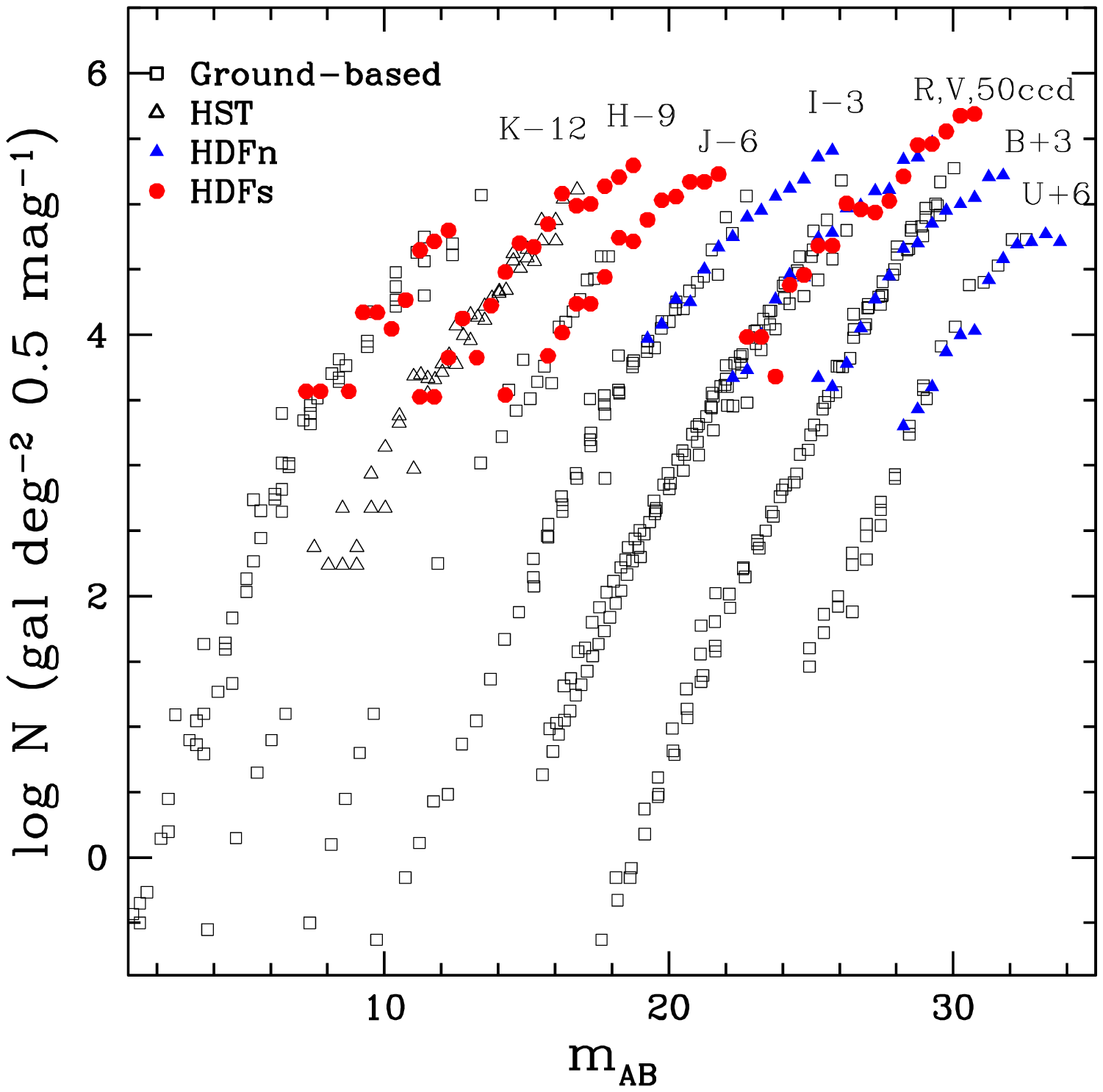,width=0.47\textwidth}}
\hfill
\parbox{0.50\textwidth}{\psfig{file=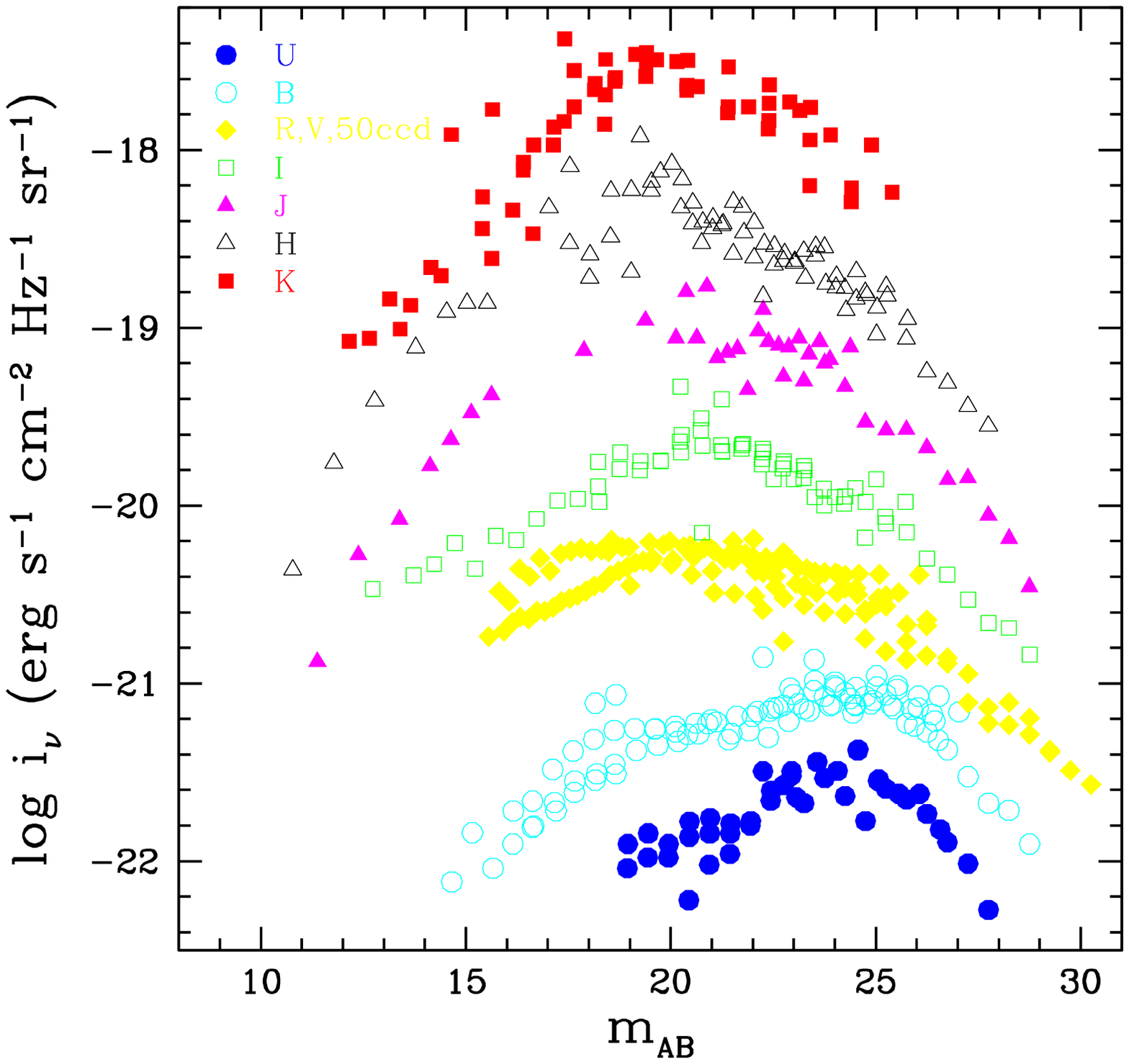,width=0.50\textwidth}}
\caption{(a)~The optical and near-infrared galaxy counts in various
passbands as a function of AB magnitude. (b)~The extragalactic
background light per magnitude bin in each passband (arbitrarily
scaled), showing the turnover and decreasing contributions at the
faintest magnitudes (see \astroph{Madau \& Pozzetti 1999}{9907315}.}
\label{fig:counts}
\end{center}
\end{figure}

\proclink{Madau} reviewed the constraints on the history of star
formation and the mass density in stars that can be derived from the
extragalactic background light (EBL). He notes that the logarithmic
slope of the faint galaxy counts drops below 0.4 at wavelengths from the
ultraviolet to the near-infrared at the limits achieved in the Hubble
Deep Fields (Figure~\ref{fig:counts}a; \astroph{Madau \& Pozzetti
1999}{9907315}). This implies the EBL is converging, with the bulk of
the light already observed (Figure~\ref{fig:counts}b), and gives a lower
limit to the EBL intensity over the range 0.2--2.2$\mu$m of
15~nW\,m$^{-2}$\,sr$^{-1}$. Combined with measurements of the
far-infrared background light from COBE, and making approximate
corrections for unresolved sources and the wings of the galaxy light
profiles, Madau estimates that the total EBL intensity is
55$\pm$20~nW\,m$^{-2}$\,sr$^{-1}$. Using stellar population models and
assumptions about the stellar initial mass function, he finds that the
EBL requires that the present-day mass density in processed gas and
stars has a lower limit of $\Omega_{g+s}h^2$$>$0.0013, corresponding to
$\langle M/L_B \rangle_{g+s}$$>$3.5. Plausible models for the
star-formation history give $\Omega_{g+s}$ twice as large. Another
constraint arising from the EBL relates to the possibility raised by the
MACHO gravitational lensing survey that galaxies' dark halos may be
composed of faint white dwarfs. Madau finds that in fact no more than
5\% of the baryons can be in form of stellar remnants without
over-producing the EBL.

Cowie and Rowan-Robinson discussed what can be learnt about galaxy
evolution by combining submillimetre observations with optical and
infrared data. \proclink{Rowan-Robinson} presented models of the
star-formation history of the universe based on results from imaging
surveys of galaxies at infrared and submillimetre wavelengths. The
European Large-Area ISO Survey
(\htmlink{ELAIS}{http://athena.ph.ic.ac.uk}; \astroph{Rowan-Robinson
et~al.\ 1999}{9906273}) has employed the
\htmlink{ISO}{http://isowww.estec.esa.nl} satellite to image
4--12~sq.deg at 6.7$\mu$m, 15$\mu$m, 90$\mu$m and 175$\mu$m, reaching
down to 3~mJy at 15$\mu$m. This survey is supported by optical and
near-infrared imaging, 21cm observations, and a
\htmlink{SCUBA}{http://www.jach.hawaii.edu/JACpublic/JCMT/scuba}
submillimetre survey of 0.02~sq.deg down to 8~mJy at 850$\mu$m.
Rowan-Robinson uses simple parametrised star-formation models to fit the
infrared and submillimetre galaxy counts (\astroph{Rowan-Robinson
1999}{9906308}). He finds that the star-formation rate increases to
$z$$\sim$1 and then remains high out to at least $z$$\sim$5. Within this
framework, the submillimetre background intensity can be used to
discriminate between cosmological models by comparing their predictions
with the observed far-infrared and submillimetre background spectrum. A
low-density ($\Omega$$\approx$0.3, $\Lambda$=0) model is preferred; a
high-density ($\Omega$=1, $\Lambda$=0) model requires a spike in the
star-formation rate at $z$$>$5, while a flat ($\Omega$$\approx$0.3,
$\Lambda$$\approx$0.7) model requires truncation of star-formation at
$z$$\ge$5.

\proclink{Cowie} noted that the submillimetre background flux indicates
that much of the energy produced by star-formation and AGN is
re-processed by dust to far-infrared wavelengths. SCUBA observations at
850$\mu$m show that the background is due to ultra-luminous infrared
galaxies at redshifts $z$$>$1. Submillimetre imaging has poor spatial
resolution, but the fact that both submillimetre and centimeter flux are
related to star-formation rate has meant that centimetre radio
observations can be used to locate the submillimetre sources
(\astroph{Barger et~al.\ 1999}{9907022}). However these objects turn out
to be extremely faint in the optical and near-infrared, and inaccessible
to spectroscopic follow-up. Nonetheless it is possible to estimate their
redshifts from the shape of their spectral energy distribution at
submillimetre and radio wavelengths. This procedure suggests they are at
redshifts $z$=1--3; if so, the submillimetre sources are a major
star-forming population at high redshift (\astroph{Cowie et~al.\
1999}{9907043}).

\proclink{Calzetti} outlined models for the evolution of the star and
gas content of galaxies, giving special consideration to the effects of
dust (\adslink{Calzetti \& Heckman 1999}{1999ApJ...519...27C}). She
estimates that the presence of dust suppresses the UV flux by a factor
of 2.5--5 at $z$=1--3, with a significant fraction of the radiation from
star-formation emerging in the far-infrared. Star formation histories
with either low star-formation rates and low dust content at high
redshift, or with high star-formation rates and high dust content, are
both compatible with the observed ultraviolet flux, but can differ by an
order of magnitude in the numbers of stars produced at $z$$>$3. Both
types of model predict that the far-infrared background is mostly
produced by high-redshift sources, but the model with a constant
star-formation rate for 1$<$$z$$<$4 predicts the dust will be hotter at
high redshift.

\section{The Future}

The `next generation' was the theme of the final session of the
conference. \proclink{Glazebrook} discussed new observational techniques
for redshift surveys. He described the application of the `nod and
shuffle' (also known as `va et vient') and `micro-slit' techniques,
which together allow near-optimal sky-subtraction at faint magnitudes
and a greatly increased multiplex gain. High-precision sky-subtraction
is achieved by simultaneously moving the electrons back and forth on the
CCD (charge-shuffling) in concert with moving the telescope back and
forth on the sky (nodding). This exact differential beam-switching
approach means that the object and the sky are observed through the same
part of the slit (or the same fibre) and with the same pixels of the
detector, but avoids the extra read-noise imposed by simply nodding
between exposures. It also offers more rapid time-sampling. With
near-perfect sky-subtraction, long slits are not needed, so one can use
small apertures (micro-slits) in the focal plane mask, and so observe
many more objects simultaneously. Glazebrook described how these
techniqes (and the introduction of a volume-phase holographic grating)
allowed the LDSS++ spectrograph on the 4-metre AAT to measure redshifts
in the Hubble Deep Field South to $R$$\approx$24 in a 12~hour exposure
(6~hours on-object). Further extensions to this technique include the
use of blocking filters to isolate small spectral ranges and so allow
more spectra to be crammed onto the detector. Alternatively, one can use
a `pseudo-slitless' mode in which the object spectra are allowed to
overlap (as in objective prism spectroscopy, but with a mask). The
ambiguity of which line belongs to which object can be resolved by
obtaining a second set of spectra with the grating rotated through
180\degree. Glazebrook outlined some case studies showing how these
techniques could significantly improve the grasp (in terms of both depth
and sample size) of the next generation of redshift surveys.

Large, deep redshift surveys, and particularly surveys of relatively
rare objects, require deep, wide-field imaging to select the target
samples.
\htmlink{[Sutherland]}{http://www.mso.anu.edu.au/DunkIsland/Proceedings/VISTA}
described the plans for a next-generation survey telescope, the 4-metre
Visible-Infrared Survey Telescope for Astronomy
(\htmlink{VISTA}{http://www-star.qmw.ac.uk/~jpe/vista}). In the visible,
VISTA will have a 1.5\degree$\times$1.5\degree\ field of view provided
by a mosaic of 50 2048$\times$4096 CCDs; in the near-infrared it will be
able to cover 1.0\degree$\times$1.0\degree\ with 4 pointings of its 9
2048$\times$2048 IR arrays. The observational strengths will be large
surveys (e.g.\ weak lensing, photometric redshifts), searches for rare
objects (e.g.\ very high redshift QSOs and galaxies, brown dwarfs) and
studies of variable objects (supernovae, microlensing, Kuiper belt
objects). VISTA will be located in Chile, possibly on Cerro Pachon, and
the current target is to achieve first light in late 2003.

\begin{figure}
\begin{center}
\hspace{0.03\textwidth}
\parbox{0.4\textwidth}{\psfig{file=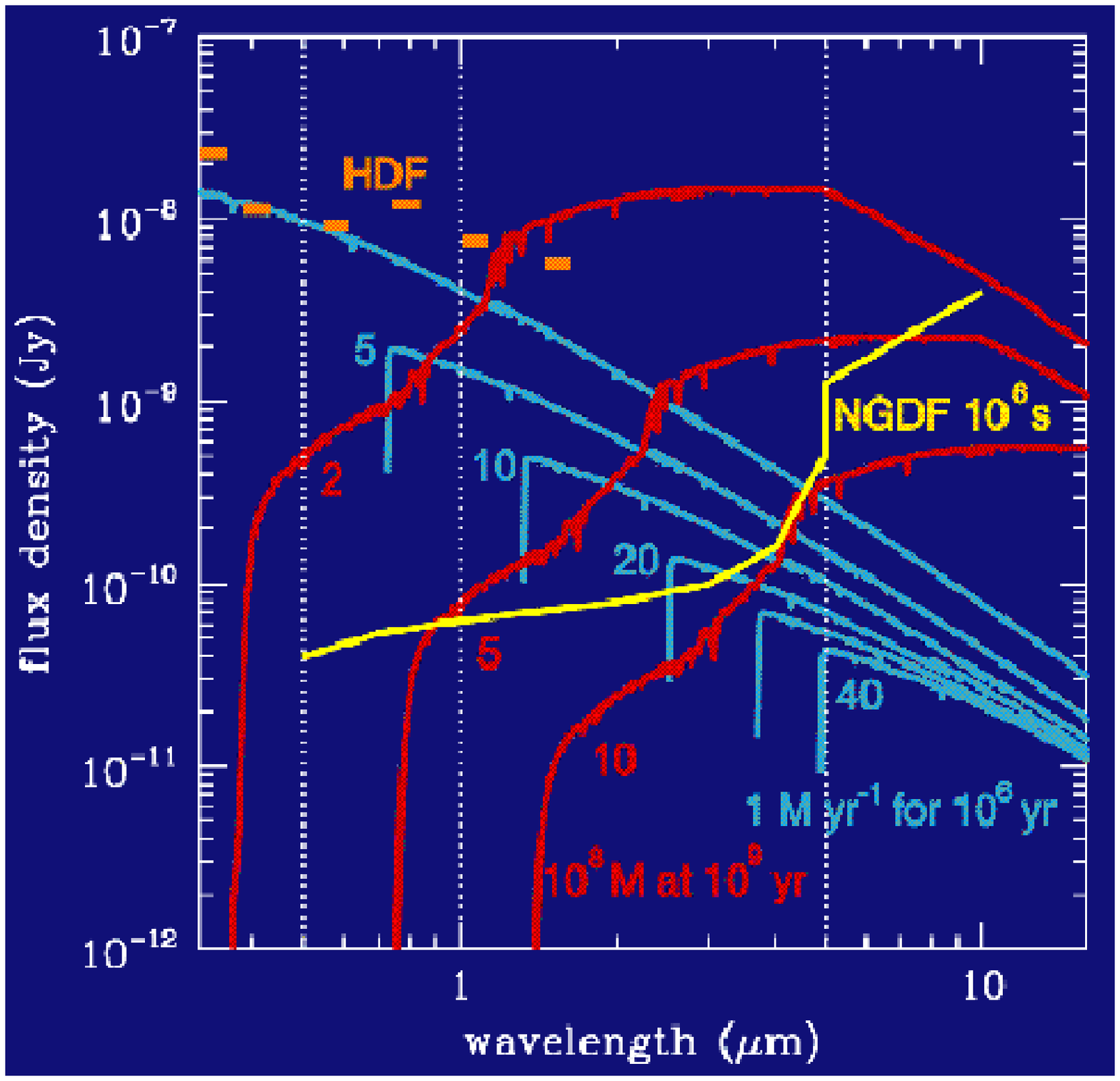,width=0.4\textwidth}}
\hfill
\parbox{0.5\textwidth}{\psfig{file=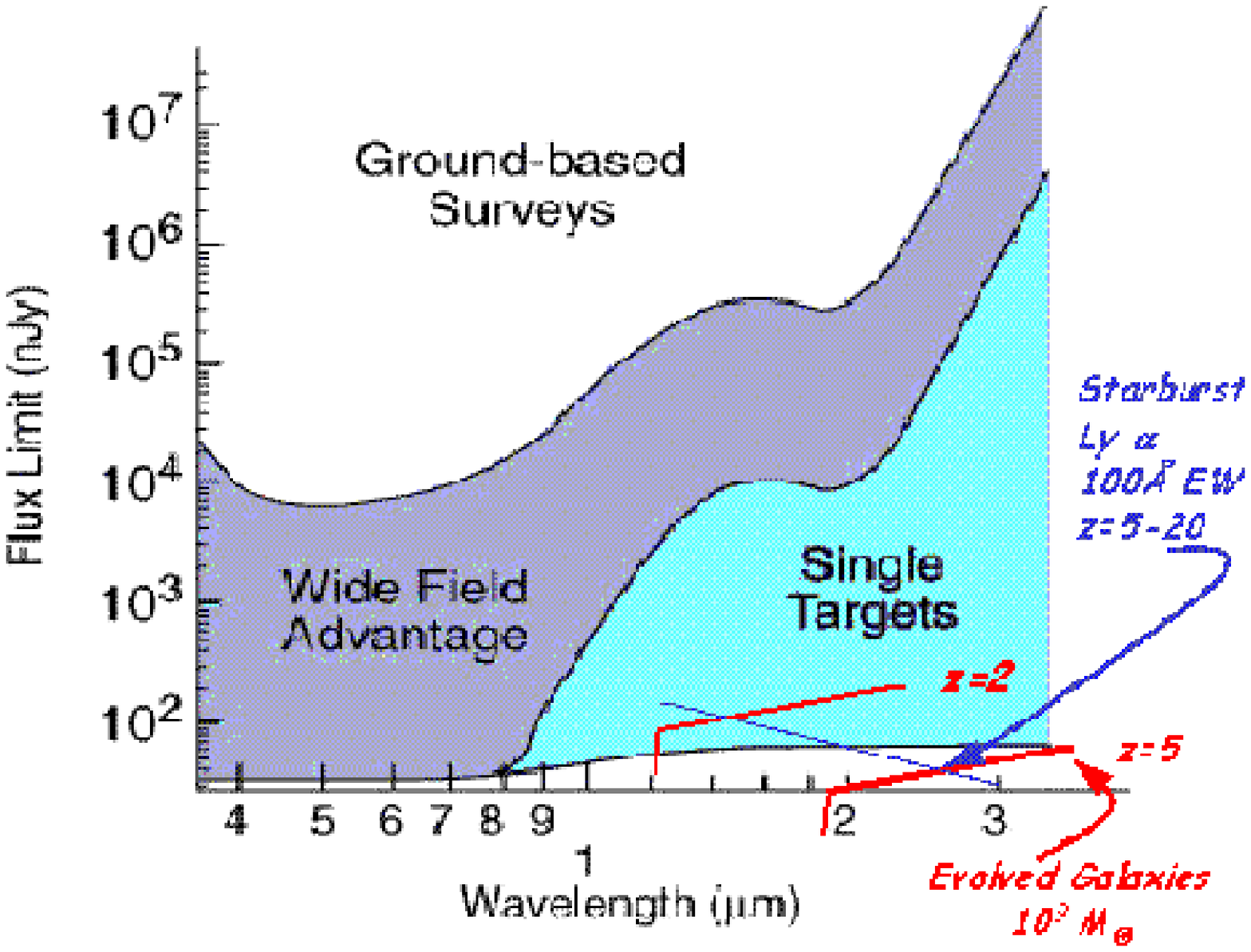,width=0.5\textwidth}}
\hspace{0.03\textwidth}
\caption{The capabilities of the Next Generation Space Telescope for
imaging and spectroscopy. (a)~The broad-band flux limits, as a function
of wavelength, reached by a proposed `Next Generation Deep Field'.
(b)~The spectroscopic flux limits, as a function of wavelength,
achievable at a resolving power of 1000 by ground-based telescopes (with
and without wide-field capability) and by NGST.}
\label{fig:ngst}
\end{center}
\end{figure}

The Next Generation Space Telescope
(\htmlink{NGST}{http://www.ngst.stsci.edu}) will be an extraordinarily
potent probe of the origin of galaxies and cosmology.
\proclink{Christian} reviewed the scientific goals and likely
configuration of NGST. The baseline plan calls for an 8-metre telescope
with wavelength coverage from 0.6--10+$\mu$m to be launched in 2008. The
\htmlink{Design Reference Mission}{http://www.ngst.stsci.edu}, which
specifies the core science goals of NGST, includes deep imaging and
spectroscopic surveys of galaxy formation and evolution. A deep imaging
survey with NGST could in principle detect a 10$^8$\Msol\ galaxy with an
old stellar population out to $z$$>$5, and a galaxy forming stars at the
rate of 1\Msol\,yr$^{-1}$ to $z$$>$10 (see Figure~\ref{fig:ngst}a). A
spectroscopic survey at a resolving power of 1000 could measure the
star-formation rate in normal galaxies at $z$$\approx$2 and in starburst
galaxies at $z$$>$5 (see Figure~\ref{fig:ngst}b). These capabilities
will allow the direct observation of the processes of star-formation in
the very early universe.

\section*{Acknowledgements}

I thank Luigi Guzzo for macros used in producing Figure~\ref{fig:onsky},
Scott Croom for some of data used in producing Figure~\ref{fig:sizevol},
John Peacock for Figure~\ref{fig:xisigpi}, the SDSS team for
Figure~\ref{fig:wkpk}, Piero Madau for Figure~\ref{fig:counts}, and
Carol Christian and Simon Lilly for Figure~\ref{fig:ngst}.

\section*{References}

\reference \astroph{Barger A.J., Cowie L.L., Richards E.A., 1999, in
Photometric Redshifts and High Redshift Galaxies, ASP Conf.\ Series, in
press}{9907022}

\reference \adslink{Boyle B.J., Croom S.M., Smith R.J., Shanks T.,
Miller L., Loaring N., 1999, in Looking Deep in the Southern Sky, eds
Morganti R., Couch W.J., Springer Verlag, p16}{1999ldss.work...16}

\reference \adslink{Calzetti D., 1997, AJ, 113,
162}{1997AJ....113..162C}

\reference \adslink{Calzetti D., Heckman T.M., 1999, ApJ, 519,
27}{1999ApJ...519...27C}

\reference \adslink{Chester T., Jarrett T., Schneider S., Skrutskie M.,
Huchra J., 1998, BAAS, 192, 5511}{1998AAS...192.5511C}

\reference \adslink{Colless M.M., 1998, in Wide Field Surveys in
Cosmology, eds Colombi S., Mellier Y., Raban B., 14th IAP Colloquium,
Editions Frontieres, p77}{1998wfsc.conf...77C}

\reference \astroph{Colless M.M., 1999, Phil. Trans. R. Soc. Lond. A,
357, 105}{9804079}

\reference \astroph{Cowie L.L., Barger A.J., 1999, in The Hy-Redshift
Universe: Galaxy Formation and Evolution at High Redshift, eds Bunker
A.J., van Breughel W.J.M., ASP Conf.\ Series, in press}{9907043}

\reference \adslink{Davis M., Faber S.M., 1998, in Wide Field Surveys in
Cosmology, eds Colombi S., Mellier Y., Raban B., 14th IAP Colloquium,
Editions Frontieres, p333}{1998wfsc.conf..333D}

\reference \adslink{Ellis R.S., Colless M.M., Broadhurst T., Heyl J.,
Glazebrook K., 1996, MNRAS, 280, 235}{1996MNRAS.280..235E}

\reference \adslink{Fan X., et al., 1999, AJ, 118,
1}{1999AJ....118....1F}

\reference \adslink{Folkes S., Ronen S., et~al., 1999, MNRAS, 308,
459}{1999MNRAS.308..459F}

\reference \adslink{Giavalisco M., Steidel C.C., Adelberger K.L.,
Dickinson M., Pettini M., Kellogg M., 1998, ApJ, 503,
543}{1998ApJ...503..543G}

\reference \adslink{Guzzo L., B\"{o}hringer H., et~al., 1999, ESO
Messenger, 95, 27}{1999Msngr..95...27G}

\reference \astroph{Kelson D.D., Illingworth G.D., van Dokkum P.G.,
Franx M., 1999, ApJ, in press}{9906152}

\reference \adslink{Kilborn K., Webster R.L., Staveley-Smith L., 1999,
PASA, 16, 8}{1999PASA...16....8K}

\reference \adslink{Koo D.C., 1998, in Wide Field Surveys in Cosmology,
eds Colombi S., Mellier Y., Raban B., Editions Frontieres,
161}{1998wfsc.conf..161K}

\reference \adslink{LeFevre O., Vettolani G., et~al., 1999, in
Observational Cosmology: The Development of Galaxy Systems, eds Giuricin
G., Mezzetti M., and Salucci P., ASP Conf.\ Series, 176,
p.250}{1999obco.conf..250L}

\reference \adslink{Loveday J., Peterson B.A., Efstathiou G., Maddox
S.J., 1992, ApJ, 390, 338}{1992ApJ...390..338L}

\reference \adslink{Loveday J., Pier J.R., 1998, in Wide Field Surveys
in Cosmology, eds Colombi S., Mellier Y., Raban B., Editions Frontieres,
317}{1998wfsc.conf..317L}

\reference \astroph{Madau P., Pozzetti L., 1999, MNRAS, in
press}{9907315}

\reference \adslink{Maddox S.J., Sutherland W.J., Efstathiou G., Loveday J.,
Peterson B.A., 1990, MNRAS, 247, 1P}{1990MNRAS.247P...1M}

\reference \astroph{Margon B., 1999, Phil. Trans. Roy. Soc. Lond. A,
357, 93}{9805314}

\reference \adslink{Mo H.J., Mao S., White S.D.M., 1999, MNRAS, 304,
175}{1999MNRAS.304..175M}

\reference \astroph{Mould J.R., Huchra J.P., et~al., 1999,
astro-ph/9909260}{9909260}

\reference \astroph{Rowan-Robinson M., Oliver S., et~al., 1999, in The
Universe as seen by ISO, eds Cox P., Kessler M.F., ESA Special
Publications Series SP-427, in press}{9906273}

\reference \astroph{Rowan-Robinson M., 1999, in Ultraluminous Galaxies:
Monsters or Babies?, eds Lutz D., Tacconi L., Astrophys.\ Space Sci., in
press}{9906308}

\reference \astroph{Sadler E.M., McIntyre V.J., Jackson C.A., Cannon
R.D., 1999, PASA, in press}{9909171}

\reference \adslink{Schmoldt I.M., Branchini E., et~al., 1999, MNRAS,
304, 893}{1999MNRAS.304..893S}

\reference \adslink{Schmoldt I.M., Saar V., et~al., 1999, AJ, 118,
1146}{1999AJ....118.1146S}

\reference \adslink{Schuecker P., B\"{o}hringer H., et~al., 1998, 19th Texas
Symposium on Relativistic Astrophysics and Cosmology, eds Paul J.,
Montmerle T., Aubourg E.}{1998tx19.confE.546S}

\reference \adslink{Strauss M., Fan X., et al., 1999, ApJ, 522,
L61}{1999ApJ...522L..61S}

\reference \adslink{Staveley-Smith L., Webster R.L., Banks G., Kilborn
K., Koribalski B., Putman M., in Looking Deep in the Southern Sky, eds
Morganti R., Couch W.J., Springer Verlag, p132}{1999ldss.work..132S}

\reference \adslink{Steidel C.C., Adelberger K.L., Giavalisco M.,
Dickinson M., Pettini M., 1999, ApJ, 519, 1}{1999ApJ...519....1S}

\reference \adslink{Sutherland W., Tadros H., et~al., 1999, MNRAS, 308,
289}{1999MNRAS.308..289S},

\reference \adslink{Tadros H., Ballinger W.E., et~al., 1999, MNRAS, 305,
527}{1999MNRAS.305..527T}

\reference \adslink{Tadros H., Efstathiou G., Dalton G., 1998, MNRAS, 296,
995}{1998MNRAS.296..995T}

\reference \adslink{van Dokkum P.G., Franx M., Fabricant D., Kelson
D.D., Illingworth G.D., 1999, ApJ, 520, 95L}{1999ApJ...520L..95V}

\reference \astroph{Vettolani G., Zucca E., et~al., 1997, A\&A, 325,
954}{9704097}

\reference \astroph{Zucca E., Zamorani G., et~al.\ 1997, A\&A, 326,
477}{9705096}

\end{document}